\documentclass[fleqn,usenatbib]{mnras}

\usepackage{newtxtext,newtxmath}

\usepackage[T1]{fontenc}

\DeclareRobustCommand{\VAN}[3]{#2}
\let\VANthebibliography\thebibliography
\def\thebibliography{\DeclareRobustCommand{\VAN}[3]{##3}\VANthebibliography}

\usepackage{graphicx}	
\usepackage{amsmath}	

\title[Spectroscopy of a lensed star at $z\simeq4.76$]{Reaching for the stars -- JWST/NIRSpec spectroscopy of a lensed star candidate at $z=4.76$}

\author[Furtak et al.]{
Lukas J. Furtak,$^{1}$\thanks{E-mail: furtak@post.bgu.ac.il}
Ashish K. Meena,$^{1}$
Erik Zackrisson,$^{2}$
Adi Zitrin,$^{1}$
Gabriel B. Brammer,$^{3}$
Dan Coe,$^{4,5,6}$
\newauthor Jos\'{e} M. Diego,$^{7}$
Jan J. Eldridge,$^{8}$
Yolanda Jim\'{e}nez-Teja,$^{9,10}$
Vasily Kokorev,$^{11}$
Massimo Ricotti,$^{12}$
\newauthor Brian Welch,$^{13,12}$
Rogier A. Windhorst,$^{14}$
Abdurro'uf,$^{6,4}$
Felipe Andrade-Santos,$^{15,16}$
Rachana Bhatawdekar,$^{17}$
\newauthor Larry D. Bradley,$^{4}$
Tom Broadhurst,$^{18,19,20}$
Wenlei Chen,$^{21}$
Christopher J. Conselice,$^{22}$
Pratika Dayal,$^{11}$
\newauthor Brenda L. Frye,$^{23}$
Seiji Fujimoto,$^{24}$
Tiger Y.-Y. Hsiao,$^{6}$
Patrick L. Kelly,$^{20}$
Guillaume Mahler,$^{25,26}$
\newauthor Nir Mandelker,$^{27}$
Colin Norman,$^{6,4}$
Masamune Oguri,$^{28,29}$
Norbert Pirzkal,$^{4}$
Marc Postman,$^{4}$
\newauthor Swara Ravindranath,$^{4}$
Eros Vanzella$^{30}$
and Stephen M.~Wilkins,$^{31,32}$
\\
$^{1}$Physics Department, Ben-Gurion University of the Negev, P.O. Box 653, Be’er-Sheva 84105, Israel\\
$^{2}$Observational Astrophysics, Department of Physics and Astronomy, Uppsala University, Box 516, SE-751 20 Uppsala, Sweden\\
$^{3}$Cosmic Dawn Center (DAWN), Niels Bohr Institute, University of Copenhagen, Jagtvej 128, K{\o}benhavn N, DK-2200, Denmark\\
$^{4}$Space Telescope Science Institute, 3700 San Martin Dr., Baltimore, MD 21218, USA\\
$^{5}$Association of Universities for Research in Astronomy (AURA) for the European Space Agency (ESA), STScI, Baltimore, MD 21218, USA\\
$^{6}$Center for Astrophysical Sciences, Department of Physics and Astronomy, The Johns Hopkins University, 3400 N Charles St. Baltimore, MD 21218, USA\\
$^{7}$Instituto de F\'isica de Cantabria (CSIC-UC). Avda. Los Castros s/n. 39005 Santander, Spain\\
$^{8}$Department of Physics, University of Auckland, Private Bag 92019, Auckland, New Zealand\\
$^{9}$Instituto de Astrof\'{i}sica de Andaluc\'{i}a, Glorieta de la Astronom\'{i}a s/n, 18008 Granada, Spain\\
$^{10}$Observat\'{o}rio Nacional - MCTI (ON), Rua Gal. Jos\'{e} Cristino 77, S\~{a}o Crist\'{o}v\~{a}o, 20921-400, Rio de Janeiro, Brazil\\
$^{11}$Kapteyn Astronomical Institute, University of Groningen, P.O. Box 800, 9700AV Groningen, The Netherlands\\
$^{12}$Department of Astronomy, University of Maryland, College Park, 20742, USA\\
$^{13}$Observational Cosmology Lab, NASA Goddard Space Flight Center, Greenbelt, MD 20771, USA\\
$^{14}$School of Earth and Space Exploration, Arizona State University, Tempe, AZ 85287-1404, USA\\
$^{15}$Department of Liberal Arts and Sciences, Berklee College of Music, 7 Haviland Street, Boston, MA 02215, USA\\
$^{16}$Center for Astrophysics, Harvard \& Smithsonian, 60 Garden Street, Cambridge, MA 02138, USA\\
$^{17}$European Space Agency (ESA), European Space Astronomy Centre (ESAC), Camino Bajo del Castillo s/n, 28692 Villanueva de la Cañada, Madrid, Spain\\
$^{18}$Department of Theoretical Physics, University of the Basque Country (UPV/EHU), Bilbao 48080, Spain\\
$^{19}$Donostia International Physics Center (DIPC), Donostia 20018, Spain\\
$^{20}$Ikerbasque (Basque Foundation for Science), Bilbao 48009, Spain\\\
$^{20}$School of Physics and Astronomy, University of Minnesota, 116 Church Street SE, Minneapolis, MN 55455, USA\\
$^{22}$Jodrell Bank Centre for Astrophysics, University of Manchester, Oxford Road, Manchester M13 9PL, UK\\
$^{23}$Department of Astronomy/Steward Observatory, University of Arizona, 933 N. Cherry Avenue, Tucson, AZ 85721, USA\\
$^{24}$Department of Astronomy, The University of Texas at Austin, Austin, TX 78712, USA\\
$^{25}$Institute for Computational Cosmology, Durham University, South Road, Durham DH1 3LE, UK\\
$^{26}$Centre for Extragalactic Astronomy, Durham University, South Road, Durham DH1 3LE, UK\\
$^{27}$Racah Institute of Physics, The Hebrew University, Jerusalem 91904 Israel\\
$^{28}$Center for Frontier Science, Chiba University, 1-33 Yayoi-cho, Inage-ku, Chiba 263-8522, Japan\\
$^{29}$Department of Physics, Graduate School of Science, Chiba University, 1-33 Yayoi-Cho, Inage-Ku, Chiba 263-8522, Japan\\
$^{30}$INAF -- OAS, Osservatorio di Astrofisica e Scienza dello Spazio di Bologna, via Gobetti 93/3, I-40129 Bologna, Italy\\
$^{31}$Astronomy Centre, University of Sussex, Falmer, Brighton BN1 9QH, UK\\
$^{32}$Institute of Space Sciences and Astronomy, University of Malta, 31 MSD 2080, Malta\\
}
\date{Accepted 2023 September 20. Received 2023 September 20; in original form 2023 July 31}

\pubyear{2023}

\begin{document}
\label{firstpage}
\pagerange{\pageref{firstpage}--\pageref{lastpage}}
\maketitle

\clearpage 

\begin{abstract}
We present JWST/NIRSpec observations of a highly magnified star candidate at a photometric redshift of $z_{\mathrm{phot}}\simeq4.8$, previously detected in JWST/NIRCam imaging of the strong lensing (SL) cluster MACS~J0647+7015 ($z=0.591$). The spectroscopic observation allows us to precisely measure the redshift of the host arc at $z_{\mathrm{spec}}=4.758\pm0.004$, and the star's spectrum displays clear Lyman- and Balmer-breaks commensurate with this redshift. A fit to the spectrum suggests a B-type super-giant star of surface temperature $T_{\mathrm{eff,B}}\simeq15\,000$\,K with either a redder F-type companion ($T_{\mathrm{eff,F}}\simeq6\,250$\,K) or significant dust attenuation ($A_V\simeq0.82$) along the line of sight. We also investigate the possibility that this object is a magnified young globular cluster rather than a single star. We show that the spectrum is in principle consistent with a star cluster, which could also accommodate the lack of flux variability between the two epochs. However, the lack of a counter image and the strong upper limit on the size of the object from lensing symmetry, $r\lesssim0.5$\,pc, could indicate that this scenario is somewhat less likely -- albeit not completely ruled out by the current data. The presented spectrum seen at a time when the Universe was only $\sim1.2$\,Gyr old showcases the ability of JWST to study early stars through extreme lensing.
\end{abstract}




\section{Introduction} \label{sec:intro}
One of the main goals of JWST \citep{gardner23}, launched in 2021 December, is to detect the very first stars and galaxies that populated the Universe. Directly observing individual stars at large extragalactic or cosmological distances is however not possible, since they would simply be too dim. Nonetheless, in the past few years several stars at cosmological distances have been detected at various redshifts, thanks to the gravitational lensing effect: individual stars in lensed background galaxies can become sufficiently magnified to be detected as they near a lensing-cluster's caustic \citep{miralda-escude91}.  

The first example of such a lensed star is \emph{Icarus} at $z\simeq1.49$ \citep{kelly18}, which was detected in the field of the galaxy cluster MACS~J1149.5+2223 as a transient event roughly atop the expected position of the critical curve, where the magnification is extremely high. A spectral energy distribution (SED) fit to the star's photometry yielded constraints on its spectral type and surface temperature \citep[a B-type super-giant with $T_{\mathrm{eff}}\simeq11\,000-14\,000$\,K;][]{kelly18}. 

Since then, several dozen other such lensed stars at cosmological distances have been detected in various strong lensing (SL) cluster fields, both serendipitously \citep[e.g.][]{kaurov19,chen19} or in designated surveys \citep[e.g. the \textit{Flashlights} survey;][]{kelly23,diego23b,meena23b}. Several were also already detected in JWST imaging campaigns during the first year of the observatory's operations \citep[e.g.][]{chen22,meena23a,diego23a,diego23c}. However, not all lensed stars necessarily appear as transient sources. Since the lens is itself made-up of small point or near-point masses such as stars, globular clusters, or black holes, a corrugated web of (micro-) caustics is formed in the source plane in the very high magnification regime around the macro-caustic. If a source star lies in this corrugated network, it can remain extremely magnified for long periods of time \citep[e.g.][]{venumadhav17,diego18,oguri18}. In general, such caustic networks represent a crucial window of opportunity for studying single stars in detail across the history of the Universe since it enables follow-up observations after detection.

Following the redshift distribution of lensed galaxies, most of the known lensed stars have been found at $z\sim1-3$. The currently farthest known lensed star is \textit{Earendel}, detected with the \textit{Hubble Space Telescope} (HST) at $z\simeq6.2$ and followed-up with JWST imaging \citep{welch22a,welch22b}. In \citet{meena23a}, we recently presented other examples at high redshift: Two lensed star candidates were detected near the symmetry point of a giant arc at $z_{\mathrm{phot}}\simeq4.8$ in the SL cluster MACS~J0647.7+7015 \citep[MACS0647; $z=0.591$;][]{ebeling07}. The arc was previously known from HST observations of the cluster \citep[e.g.][]{zitrin11a,zitrin15a,postman12,coe13}. The brighter and more reliable of these two lensed star candidates, MACS0647-star-1, had been classified from broad-band photometry in \citet{meena23a} to be, most probably, a B-type star with $T_{\mathrm{eff}}\sim10\,000$\,K and $M\gtrsim20\,\mathrm{M}_{\odot}$. It has now been observed spectroscopically with JWST, delivering an unprecedented spectrum of a lensed star at such a cosmological distance\footnote{At lower-redshifts, a ground-based spectrum was taken for the lensed star \textit{Godzilla} \citep[$z=2.37$;][]{vanzella20,diego22} and a JWST/NIRSpec spectrum of \textit{Earendel} is also expected soon (Welch et al. in prep.)}, showcasing JWST's ability to directly probe early stars through extreme lensing. 

In this letter, we present the spectrum of MACS0647-star-1, and analyze it to derive the star's properties. This work is organized as follows: In \S \ref{sec:data} we describe the data and observations used in this work. In \S \ref{sec:spectroscpy} we analyze the spectrum, and the results are discussed in \S \ref{sec:SED-modeling}. The work is concluded in \S \ref{sec:conclusion}. Throughout, we assume a standard flat $\Lambda$CDM cosmology with $H_0=70\,\frac{\mathrm{km}}{\mathrm{s}\,\mathrm{Mpc}}$, $\Omega_{\Lambda}=0.7$, and $\Omega_\mathrm{m}=0.3$. Magnitudes are quoted in the AB system \citep{oke83} and uncertainties represent $1\sigma$ ranges unless stated otherwise.

\section{Data} \label{sec:data}
\begin{figure}
    \centering
    \includegraphics[scale=0.68]{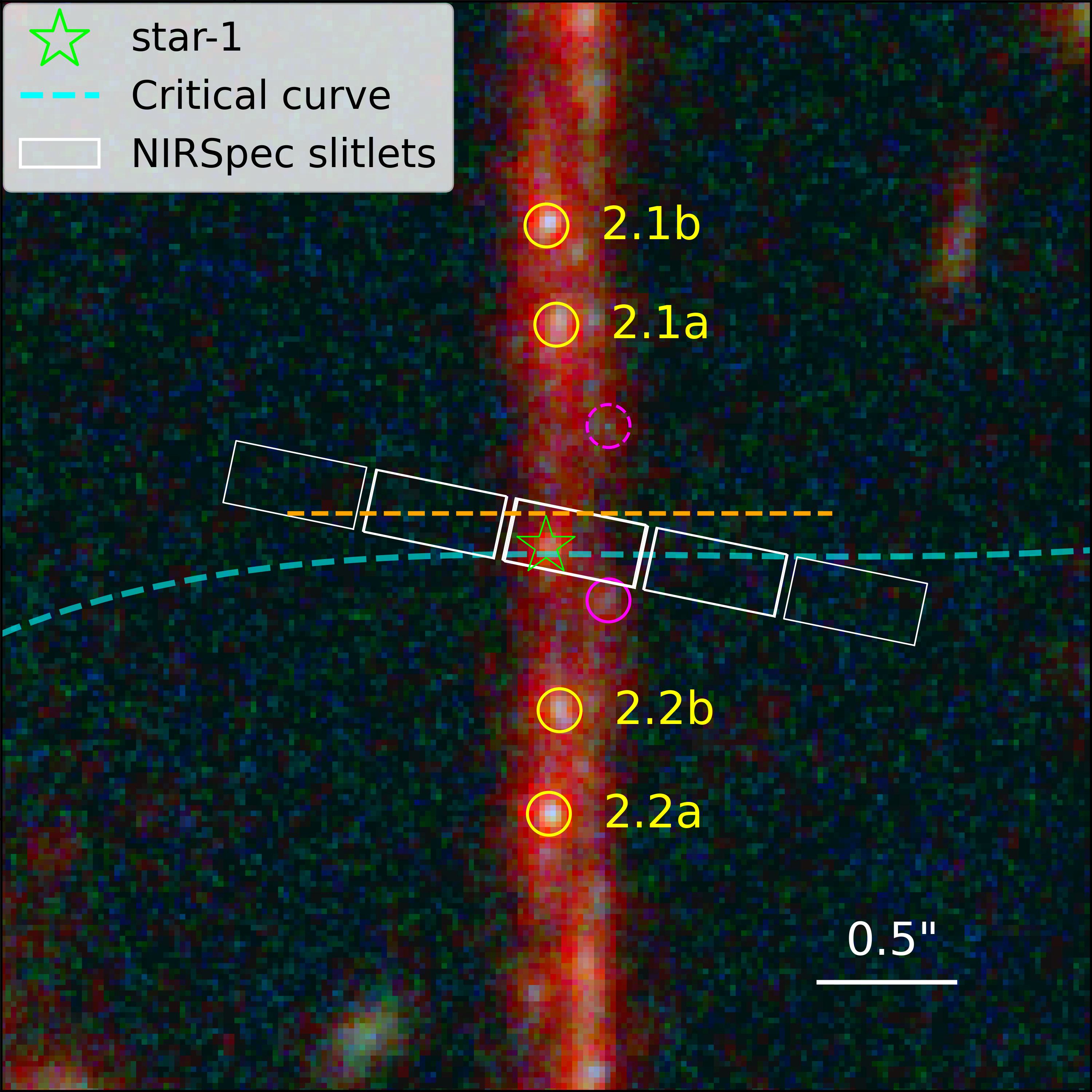}
    \caption{JWST/NIRCam composite-color image (Blue: F115W+F150W, Green: F150W+F200W, Red: F277W+F356W+F444W) of the $z=4.758$ caustic-crossing arc in MACS0647 and the lensed stars that it hosts. MACS0647-star-1 is highlighted with a green star and we also show the location of MACS0647-star-2 (magenta solid circle) even though it was not included in the NIRSpec slitlets (white). We also mark a possible counter image for star-2 in dashed purple. The critical curve of the \citet{meena23a} SL model of the cluster is overlaid in dashed turquoise and multiply-imaged clumps within the arc are highlighted in yellow. The symmetry line between the multiply-imaged clumps is shown in dashed orange.}
    \label{fig:slit}
\end{figure}

The data used in this work were taken with JWST in the framework of GO~program 1433 (PI: D.~Coe). The spectrum of MACS0647-star-1 was taken with the \textit{Near Infrared Spectrograph} \citep[NIRSpec;][]{jakobsen22,boeker23} aboard JWST in its multi-object spectroscopy (MOS) mode \citep{ferruit22} using the micro-shutter array (MSA) in standard 3-slitlet nods, to allow local sky subtraction, with a total exposure time of 1.8\,h. The slit position and orientation can be seen in Fig.~\ref{fig:slit}. The resulting prism spectrum achieves resolutions of $R\sim30-300$ in the wavelength range from $0.6-5.3\,\mu$m. Additional details regarding the observations, data reduction with the STScI JWST pipeline\footnote{\url{https://github.com/spacetelescope/jwst}}, and spectral extraction with \texttt{MSAEXP}\footnote{\url{https://github.com/gbrammer/msaexp}} are described in detail in \citet{hsiao23b}.

The imaging data for MACS0647 are comprised of deep \textit{Near Infrared Camera} \citep[NIRCam;][]{rieke23} observations in 7 filters: F115W, F150W, F200W, F277W, F365W, F444W, and F480M. The F480M imaging and a second epoch of F200W-band exposure were taken at the same time as the NIRSpec observations described above. The data were reduced using the \texttt{grism redshift and line analysis software for space-based spectroscopy} pipeline \citep[\texttt{grizli};][]{grizli22} and achieve exposure times of 35\,minutes per band. We refer the reader to \citet{hsiao23a} and \citet{meena23a} for the details regarding the imaging data and their reduction, and the photometric analysis of MACS0647-star-1.

\section{Spectroscopic analysis} \label{sec:spectroscpy}

\begin{figure*}
    \centering
    \includegraphics[width=\textwidth]{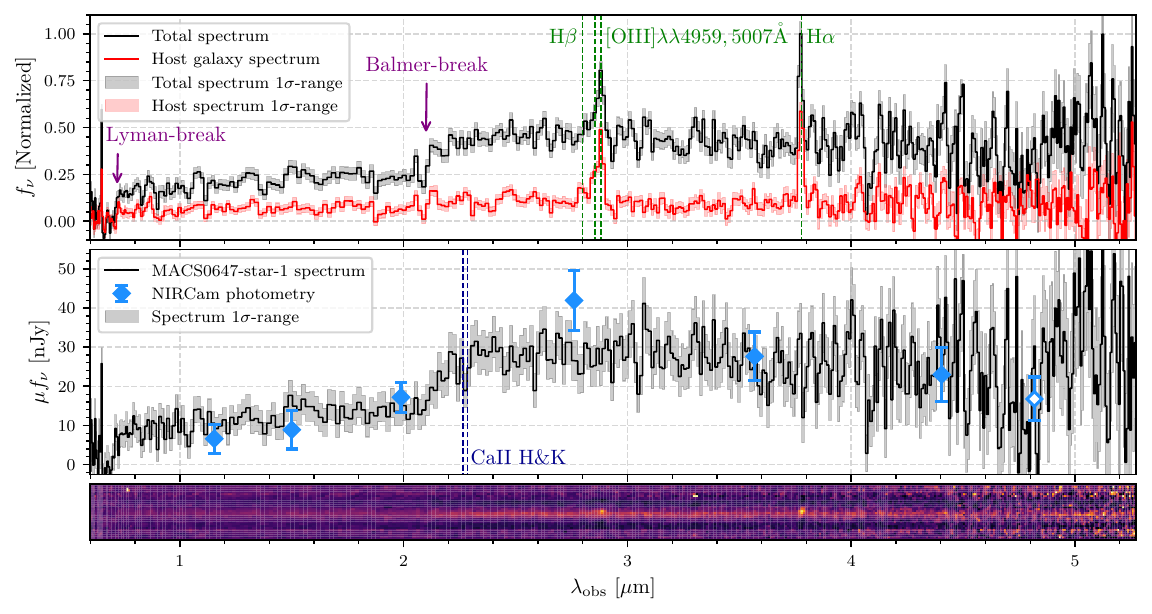}
    \caption{JWST/NIRSpec spectroscopy of MACS0647-star-1. \textit{Bottom}: 2D-spectrum in the JWST/NIRSpec slitlet shown in Fig.~\ref{fig:slit}. The two emission lines are spatially slightly offset from the continuum in the central trace, suggesting they originate from the host galaxy rather than the lensed star itself. \textit{Top}: Total spectrum extracted by \texttt{MSAEXP} (black) and the host galaxy spectrum (red), in arbitrary flux units. The host spectrum is extracted in a smaller box, centered vertically on the offset emission lines' position instead of the continuum. These two prominent emission lines (green) are identified as the (unresolved) [\ion{O}{iii}]$\lambda\lambda4959,5007$\AA-doublet and H$\alpha$ and thus yield a spectroscopic redshift measurement of $z=4.758\pm0.004$. \textit{Middle}: Host-subtracted and flux-calibrated NIRSpec spectrum of MACS0647-star-1 (black) and the corresponding NIRCam photometry (blue squares; the open square represents second-epoch photometry). The spectrum displays clearly visible Lyman- and Balmer-breaks but, as expected given the signal-to-noise, no significant absorption features. That being said, we tentatively observe slight troughs where the \ion{Ca}{ii} H and K lines (dark blue) can be expected given the redshift. The spectrum as shown here is not corrected for gravitational magnification.}
    \label{fig:spectrum}
\end{figure*}

The MSA slitlets, shown in Fig.~\ref{fig:slit}, contain both the emission of the star and some diffuse emission from the host galaxy around it. The spectrum, which can be seen in Fig.~\ref{fig:spectrum}, shows two emission lines that are slightly offset (by 2 pixels) on the spatial axis from the central continuum trace. This means that they originate from a different location than the star, i.e. most probably the host galaxy. Together with a prominent Balmer-break observed at $\lambda\sim2.1\,\mu$m and a weaker Lyman-break at $\lambda\sim0.7$\,$\mu$m, we measure the redshift of the host by identifying these two emission lines as the [\ion{O}{iii}]$\lambda\lambda4959,5007$\AA-doublet and H$\alpha$ at $z\simeq4.8$. There is also a tentative detection of H$\beta$\footnote{Assuming a \citet{calzetti00} law, this results in a tentative measurement of $A_V=1.67\pm0.33$ for the host galaxy based on the Balmer-decrement \citep[][]{dominguez13}. Note that this does not affect our spectral modeling results in section~\ref{sec:SED-modeling} though since that is only sensitive to very localized dust in the source's immediate surroundings.}. The line centers are determined by fitting a Gaussian to each continuum-subtracted emission line with \texttt{specutils} \citep[\texttt{v1.10.0};][]{specutils23} and then estimating the uncertainties with a Monte-Carlo Markov Chain (MCMC) analysis using \texttt{emcee} \citep{foreman-mackey13}. This yields a spectroscopic redshift measurement of $z_{\mathrm{spec}}=4.758\pm0.004$ for this arc, confirming the photometric redshift of $z_{\mathrm{phot}}=4.79_{-0.15}^{+0.07}$ from \citet{meena23a}.

To separate the star candidate's spectrum from contamination by the host arc, we use the fact that the emission lines are spatially offset from the central continuum trace to extract a host galaxy spectrum. This is done by taking a smaller box above the continuum but centered on the emission lines' vertical position, using the same algorithm as the \texttt{MSAEXP} pipeline (see section~\ref{sec:data}). The thus obtained 1D-spectrum of the host is then weighted by its contribution to the total trace fit by \texttt{MSAEXP} in the initial data reduction and extraction (see e.g. \citealt{hsiao23b}). As can be seen in the top panel of Fig.~\ref{fig:spectrum}, the host galaxy spectrum clearly shows the emission lines mentioned above and a low level of continuum emission but lacks a strong Balmer-break (red-to-blue flux ratio of $1.6\pm0.1$ measured between the two sides of the break). This host spectrum is then simply subtracted from the total spectrum extracted by \texttt{MSAEXP} to obtain the spectrum of the star. Since the signals of the host and the star are inevitably merged together in the NIRSpec spectrum, a perfect host subtraction can probably not be achieved. This means that we are probably over- or under-subtracting the host contribution. Since the host continuum is fairly flat (see top panel of Fig.~\ref{fig:spectrum}), we do not expect this to affect the shape of the spectrum though, only its total flux level. Because of that, we re-normalize the host-subtracted spectrum to the six bands of broad-band photometry of MACS0647-star-1 measured in \citet{meena23a}, which also effectively corrects for slit losses. While some fluctuations in magnification can be expected for a persistent lensed star near the caustic \citep[e.g.][]{welch22a,jimenez-vicente22}, the F200W imaging taken at the same time as the NIRSpec observations (see section~\ref{sec:data}) does not show any flux evolution between the two epochs: i.e. $28.31\pm0.24$ and $28.27\pm0.22$ magnitudes in the first and second epochs respectively, measured with \texttt{CICLE} \citep{jimenez-teja16} as in \citet{meena23a}. We note that given its position, there is a slight possibility that some residue flux from MACS0647-star-2 (see Fig.~\ref{fig:slit}) entered the slit as well. That contribution would be included in our `host' spectrum and therefore be subtracted out though.

\begin{figure*}
    \centering
    \includegraphics[width=\textwidth]{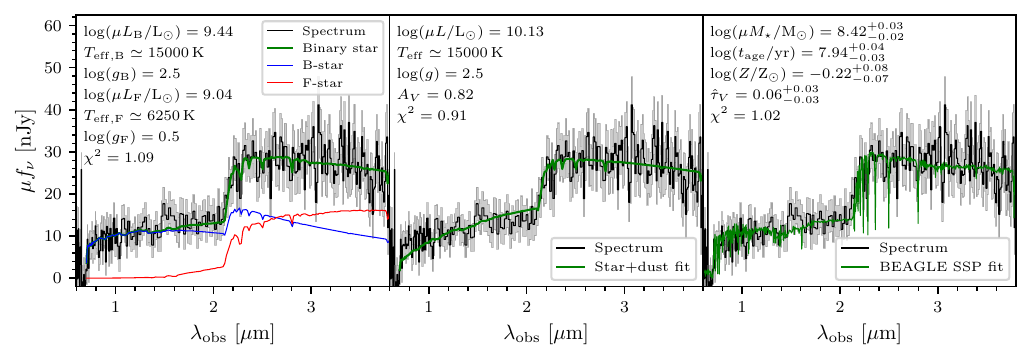}
    \caption{Spectral fits of star and star-cluster templates to the spectrum of MACS0647-star-1, which is shown in black. \textit{Left}: A two-star fit (green). The more luminous component, a B-type star, is shown in blue and the fainter, an F-type star, in red. \textit{Middle}: A single star fit with significant dust attenuation. \textit{Right}: Maximum-a-posteriori (MAP) solution for a globular cluster-type (SSP) fit with \texttt{BEAGLE} (green).}
    \label{fig:SED-fits}
\end{figure*}

The thus cleaned and flux-calibrated spectrum of the lensed star is shown in the middle panel of Fig.~\ref{fig:spectrum}. The host subtraction described above effectively removed the emission lines, indicating a fair subtraction. While the spectrum matches the NIRCam photometry in all bands within the $1\sigma$-range, the Balmer-break seems shallower than that estimated from the photometry in \citet{meena23a}. While we do not observe any significantly strong absorption features, which is in agreement with the expected signal-to-noise ratio (SNR), the spectrum shows tentative troughs ($\sim1\sigma$) at the expected wavelengths of the \ion{Ca}{ii} H and K lines, and more marginally, also near the Fraunhofer G-band, H$\gamma$ and H$\beta$ wavelengths (see also Fig.~\ref{fig:SED-fits}). Given the SNR, however, at this point we should acknowledge that this may simply be coincidental, and that deeper observations will be needed to probe if these are real. 

\section{Spectral modeling} \label{sec:SED-modeling}

In order to determine the properties of the source, a more quantitative analysis of the spectrum presented in section~\ref{sec:spectroscpy} is required. We therefore perform detailed spectral modeling, both considering the source to be one, or two individual stars (section~\ref{sec:star-SED}), or a globular cluster-like object bound to the host galaxy (section~\ref{sec:GC-SED}) and discuss the results in section~\ref{sec:size_discussion}.

\subsection{Stellar modeling} \label{sec:star-SED}
The Balmer-break of the spectrum is weaker than inferred from the photometry (see Fig.~\ref{fig:spectrum}), such that the simple single-star solution from \citet{meena23a} should be revised here with the spectroscopic data. We therefore attempt to fit the spectrum of MACS0647-star-1 with either a single-star SED with dust attenuation or as a superposition of two single-star SEDs without attenuation. All SED templates are based on the \citet{lejeune97} set of stellar atmosphere spectra redshifted to $z=4.76$ and re-sampled to the observed resolution. At the spectral resolution of the NIRSpec prism, the effect of metallicity is negligible on the shapes of these SEDs, and we therefore only present fits for metallicity $[\mathrm{M}/\mathrm{H}]=-1$ (but have verified that $[\mathrm{M}/\mathrm{H}]=0$ does not have any significant impact on our conclusions). Since we are likely dealing with very massive stars, we moreover limit the fits to the lowest surface gravities available at each effective temperature $T_\mathrm{eff}$ within the grid. For the dust attenuation, we adopt the \citet{calzetti00} law, as parameterized by \citet{Li08}. 

The resulting best-fit SEDs are shown in the left-hand and middle panels of Fig.~\ref{fig:SED-fits}. In the two-star case, the primary (i.e more luminous) component is a B-type star of surface temperature $T_{\mathrm{eff,B}}\simeq15\,000$\,K and (magnified) luminosity $\log(\mu L_{\mathrm{B}}/\mathrm{L}_{\odot})\simeq9.43$ with an F-type star companion of $T_{\mathrm{eff,F}}\simeq6\,250$\,K and $\log(\mu L_{\mathrm{F}}/\mathrm{L}_{\odot})\simeq9.04$. Under reasonable assumptions on the magnification, both components correspond to high-mass stars (initial masses $\gtrsim10\,\mathrm{M}_{\odot}$ for $\mu<50\,000$), with $\mu\gtrsim1\,300$ required to bring the more luminous star down to an initial mass below $\lesssim100\,\mathrm{M}_{\odot}$. This combination of stars would be unexpected for several reasons. Unless $\mu>1\,000-3\,000$, both stars violate the empirical Humphreys-Davidson luminosity limit \citep[caused by envelope-stripping due to stellar winds][]{humphreys79}. Moreover, the recent stellar evolutionary models for high-mass stars by \citet{szecsi22} suggest that the evolutionary states corresponding to the best-fitting surface temperature would be extremely brief ($\lesssim10^5$\,yr), except at very specific masses and metallicities. Under the assumption that the two stars have similar magnifications, the ratio of bolometric luminosities ($L_{\mathrm{B}}/L_{\mathrm{F}}\approx2.4$) also causes a potential age problem. While our constraints on the source size in no way require the two stars to be part of the same binary system, they are still likely to be young members of the same star cluster, and hence have similar ages. In the context of single-star evolution, stars with higher luminosity and initial mass typically evolve to low-$T_{\mathrm{eff}}$ states ahead of their lower-$L$ and less massive counterparts, whereas the opposite seems to be the case here. While blue loops along the evolutionary track of the more massive star could potentially provide an explanation, this requires finely tuned ages given the large $L_{\mathrm{B}}/L_{\mathrm{F}}$-ratio. This problem could also be resolved in a scenario where the inferred luminosity ratio is strongly affected by dust attenuation along the line of sight, either from circum-stellar or interstellar dust. In a situation where the light from the lower-$T_{\mathrm{eff}}$ star is experiencing significantly more dust attenuation than the higher-$T_{\mathrm{eff}}$ star, the intrinsic $L_{\mathrm{B}}/L_{\mathrm{F}}$-ratio would be lower than that inferred from our current fit. We have also searched through the suite of binary models from the \textit{Binary Population And Spectral Synthesis} (BPASS; v2.2) results \citep{stanway18}. The only matches with the effective temperatures and luminosity ratios matching our two-star fit are low-mass stars of a few $\mathrm{M}_{\odot}$. These would require even more extreme magnifications though due to the relatively low luminosities of low-mass stars. This suggests that our object is unlikely to be part of a co-evolving binary. One more possibility that needs mentioning is a redder star accompanied by a stellar-mass black hole accretion disc as suggested in \citet{windhorst18}, although that would have much higher temperatures and therefore a bluer spectrum than our object.

In the single-star scenario, we are also looking at a B-type star of $T_{\mathrm{eff}}\simeq15\,000$\,K which is however somewhat brighter ($\log(\mu L/\mathrm{L}_{\odot})\simeq10.13$) and considerably reddened by dust to $A_V\simeq0.82$\,magnitudes. In this case, a magnification of $\mu\gtrsim7\,000$ would be required to bring the star below $\lesssim100\,\mathrm{M}_{\odot}$, based on the evolutionary tracks of \citet{szecsi22}. In both cases the fit is driven by a $T_{\mathrm{eff}}\simeq15\,000$\,K B-type super-giant and the companion or dust attenuation are reducing the strength of the Balmer-break and flattening the continuum at longer wavelengths to match the spectrum. The surface temperature is somewhat higher than inferred from the photometry in \citet{meena23a} ($T_{\mathrm{eff}}\simeq10\,500$\,K). The present data do not enable us to clearly distinguish between these two solutions as both scenarios provide a good fit to the spectrum ($\chi^2\simeq1$). The single-star solution is easier to explain using current stellar evolutionary models however.

\subsection{Stellar population modeling} \label{sec:GC-SED}
In order to explore the possibility of our source being a star-cluster instead of one or two single lensed stars, we perform an additional spectral fit with the \texttt{BayEsian Analysis of GaLaxy sEds} tool \citep[\texttt{BEAGLE};][]{chevallard16}. We use the latest version of the \citet{bc03} stellar population models combined with nebular emission models by \citet{gutkin16}, a \citet{charlot00} dust attenuation law and the \citet{inoue14} intergalactic medium (IGM) attenuation models. The star cluster is parametrized as a single stellar population (SSP) with four free parameters: the (magnified) stellar mass ($\log(\mu M_{\star}/\mathrm{M}_{\odot})\in[4,10]$), age ($\log(t_{\mathrm{age}}/\mathrm{yr})\in[6,t_{\mathrm{Universe}}]$), dust attenuation ($\hat{\tau}_V\in[0,0.1]$) and metallicity ($\log(Z/\mathrm{Z}_{\odot})\in[-3,0.4]$). The result of this fit is shown in the right-hand panel of Fig.~\ref{fig:SED-fits}.

\subsection{A star or a star cluster? Insight from lensing} \label{sec:size_discussion}
All three options shown in Fig.~\ref{fig:SED-fits}, namely a single star+dust, two stars, or a star cluster, yield similarly good fits to the spectral data, rendering all options viable. However, we can use constraints from lensing symmetry, magnification, and possible flux variability, to obtain some more insight regarding the nature of the source. Indeed, given that the source lies very close to the critical curves where the magnification gets extremely high, the latter is very difficult to measure. We start by noting that based on the SL model by \citet{meena23a}, the total magnification at the position of star-1, which lies $\sim$0.04\arcsec from the critical curves of the model, is $\mu\sim600$, whereas the tangential magnification is $\mu_{\mathrm{t}}\sim500$. Given that the lensed source remains point-like, we can adopt the point-spread-function (PSF) of 0.04\arcsec, which translates into an upper limit of $r\lesssim0.3\,\mathrm{pc}$ on the size of the source. In fact, since the position of the critical line can be constrained independently from the SL model by symmetry considerations \citep[Fig.~\ref{fig:slit}; see also Figure~3 in][]{meena23a}, we can also infer a more model-independent estimate. Star-1 lies about $0.1\arcsec$ from the deduced symmetry point, which results in an upper-limit of $r\lesssim0.5\,\mathrm{pc}$, corresponding to a magnification of $\mu\sim250$.

Note that if this object were a star cluster, then from the stellar population fit in section~\ref{sec:GC-SED} it would have a magnified stellar mass of $\log(\mu M_{\star}/\mathrm{M}_{\odot})=8.42_{-0.02}^{+0.03}$. Combining this stellar mass with the magnification and size estimates, we obtain that the intrinsic (i.e. de-lensed) projected stellar mass density would be $\Sigma_{\star}\simeq2.1\times10^6\,\mathrm{M}_{\odot}/\mathrm{pc}^2$. This is more than an order of magnitude higher than the highest densities observed in globular clusters \citep[e.g.][]{norris14} or young massive star clusters known to date, both at low and high redshifts \citep[e.g.][]{brown21,vanzella23}, and two orders of magnitude higher than the highest densities found in high-resolution simulations \citep[e.g.][]{he19,garcia23}, therefore making the star-cluster option more challenging. 

The star-cluster option, on the other hand, is consistent with the lack of flux variability in the F200W-band which repeated in both epochs: since star clusters are larger and much more complex compared to stars, only very weak fluctuations are expected \citep[e.g.][]{dai21}. Nevertheless, stars in the corrugated network of micro-caustics which forms around the macro-caustic are also expected to show only mild fluctuations \citep[e.g.][]{welch22a}. For dense networks of micro-caustics, the time variation can be very rapid but small in amplitude and thus the source may appear constant for a relatively long period of time. The mere two epochs in one band are likely insufficient for drawing a conclusion based on variability at this stage.  

There is another challenge to the star-cluster possibility: the source lacks a counter image on the other side of the critical curves (or the symmetry axis). For the source to be a star-cluster and not show a counter image, one would have to invoke a milli-lens to create a local area of de-magnification and thus `hide' the counter image. While plausible, this would require some fine-tuning, which complicates the interpretation. In that regard it is worth mentioning another possibility: In the case of ``wave dark matter'', the network of caustics is pervasive and thus small sources can be strongly magnified without new counter images appearing \citep{amruth23}. In this framework the source could in principle be a star cluster, without a need to invoke additional perturbers.

The most plausible solution may simply be, as originally proposed, a star very near the macro-caustic, i.e in the corrugated network of micro-caustics, in which case no counter images are necessarily expected on the other side of the macro critical curves. As is evident in Fig.~\ref{fig:slit}, the star lies between the model's critical curve and the model-independent symmetry line. The true position of the critical curve lies most likely somewhere between these two lines, meaning that star-1 is $\lesssim0.1\arcsec$ from it. For comparison, the size of the corrugated network, which depends on the micro-lens surface density and its mass function, as well as on the underlying lensing quantities (such as the lens-mapping Jacobian's eigenvalues; \citealt{venumadhav17}), is of the same order -- assuming that about 1\% of the surface mass density consists of microlenses. Note also that in order for an object to remain a point-source in such a scenario, i.e. to appear smaller than the given PSF of 0.04\arcsec, its radius cannot exceed $r\lesssim0.01$\,pc, corresponding to a magnification of $\mu\sim\mathcal{O}(10^{4})$ \citep[see formulae in][]{venumadhav17,meena23a}. 

Based on the above considerations we therefore conclude that the scenario of a young globular cluster is somewhat less likely, even though it cannot be ruled out with the present data either and remains a viable option. Distinguishing better the two scenarios would require a higher SNR which would allow us to measure absorption lines, as well as additional visits to examine further the flux variability.

\section{Conclusion} \label{sec:conclusion}
We have presented a JWST/NIRSpec prism spectrum of the extremely lensed star candidate MACS0647-star-1, detected previously in JWST/NIRCam imaging of the SL cluster MACS~J0647.7+7015. The slit covers also a portion of the host arc which shows two prominent emission lines, identified as the unresolved [\ion{O}{iii}]$\lambda\lambda4959,5007$\AA-doublet and the H$\alpha$ line. These allow us to precisely measure the spectroscopic redshift at $z_{\mathrm{spec}}=4.758\pm0.004$. We subtract the contribution of the host galaxy to obtain a clean spectrum of the star, which shows clear Lyman- and Balmer-breaks at the same redshift but -- given the SNR -- lacks significant absorption (or emission) features. Spectral fits considering both a single and a two-star solution find this object to most likely be a B-type super-giant star of surface temperature $T_{\mathrm{eff}}\simeq15\,000$\,K with either an F-type companion of $T_{\mathrm{eff}}\simeq6\,250$\,K or significant dust attenuation of $A_V\simeq0.82$ originating from the star's surroundings. We also explore the possibility of the object being a globular cluster progenitor and fit it with an SSP template which results in a (magnified) stellar mass estimate of $\mu M_{\star}\simeq10^{8.4}\,\mathrm{M}_{\odot}$. Given that this would result in an extremely high stellar mass density when combined with the lensing constraints on the radius of our object ($r\lesssim0.5$\,pc, for $\mu\gtrsim250$) and would produce a counter image which we do not detect, this solution is somewhat disfavored compared to the star solutions.

By delivering JWST's first spectrum of an extremely magnified star candidate at nearly $z\sim5$, this work clearly showcases the phenomenal potential of JWST to probe individual stars across the history of the Universe with the aid of gravitational lensing. In order to explore the properties of these sources in detail and clearly distinguish between a single star and a star-cluster, various visits are needed to examine variability, as well as deeper observations to follow-up on the tentative absorption lines -- and at potentially higher spectral resolutions than provided by the NIRSpec prism mode. These might be achieved, for example, with the higher-resolution gratings on NIRSpec (optimally with spectral resolutions of $R\sim1\,000$), but would require longer exposure times and, preferably, brighter stars. 

Some prominent lensed stars have also already been targeted spectroscopically from the ground at lower redshifts (e.g. \textit{Godzilla}). Obtaining spectra of lensed stars could thus represent a compelling science case also for upcoming extremely large telescopes, combining unprecedented spatial resolution and the collecting power of 30-40\,m class mirrors with significantly higher spectral resolution ($R\gtrsim10\,000$ class) spectrographs.

\section*{Acknowledgements}
We dedicate this paper to our colleague Mario Nonino who passed away during the course of this work. Mario was a very capable, curious, and enthusiastic astronomer. He was always happy to help advance projects, reduce, and analyze data -- even while leading large programs of his own. Above all, Mario was an amiable colleague and a good friend to many. He will be dearly missed.

L.F. would like to thank Lisa W\"{o}lfer and Ivo Labb\'{e} for useful discussions.

The BGU group acknowledges support by Grant No. 2020750 from the United States-Israel Binational Science Foundation (BSF) and Grant No. 2109066 from the United States National Science Foundation (NSF), and by the Ministry of Science \& Technology of Israel. E.Z. acknowledges project grant 2022-03804 from the Swedish Research Council (Vetenskapsr\aa{}det) and funding from the Swedish National Space Board. Y.J-T. acknowledges financial support from the European Union’s Horizon 2020 research and innovation program under the Marie Sk\l{}odowska-Curie grant agreement No.~898633, the MSCA IF Extensions Program of the Spanish National Research Council (CSIC), and the State Agency for Research of the Spanish MCIU through the Center of Excellence Severo Ochoa award to the Instituto de Astrof\'{i}sica de Andaluc\'{i}a (SEV-2017-0709). PD acknowledges support from the NWO grant 016.VIDI.189.162 (``ODIN") and from the European Commission's and University of Groningen's CO-FUND Rosalind Franklin program. TH and A were funded by a grant for JWST-GO-01433 awarded by STScI. EV acknowledges financial support through grants PRINMIUR~2017WSCC32, 2020SKSTHZ and INAF ``main-stream'' grants~1.05.01.86.20 and~1.05.01.86.31. EV further acknowledges support from the INAF GO Grant 2022. This project has received funding from NASA through the NASA Hubble Fellowship grant HST-HF2-51505.001-A awarded by the STScI.

This research made use of \texttt{Astropy},\footnote{\url{http://www.astropy.org}} a community-developed core Python package for Astronomy \citep{astropy13,astropy18} as well as the packages \texttt{NumPy} \citep{vanderwalt11}, \texttt{SciPy} \citep{virtanen20}, \texttt{Matplotlib} \citep{hunter07} and the \texttt{MAAT} Astronomy and Astrophysics tools for \texttt{MATLAB} \citep[][]{maat14}.


\section*{Data Availability}
The NASA/ESA/CSA JWST data used in this work are publicly available on the \texttt{Barbara A. Mikulski Archive for Space Telescopes} (\texttt{MAST}) at the \textit{Space Telescope Science Institute} (STScI) under program ID JWST GO~1433. STScI is operated by the Association of Universities for Research in Astronomy, Inc. under NASA contract NAS~5-26555. Reduced data products, catalogs and lens models are also publicly available on the Cosmic Spring website\footnote{\url{https://cosmic-spring.github.io/index.html}} and the \texttt{Dawn JWST Archive} (\texttt{DJA})\footnote{\url{https://dawn-cph.github.io/dja/index.html}}. DJA is an initiative of the Cosmic Dawn Center, which is funded by the Danish National Research Foundation under grant No.~140.



\bibliographystyle{mnras}
\bibliography{references} 



\appendix


\bsp	
\label{lastpage}
\end{document}